\documentstyle[prl,aps]{revtex}
\draft
\begin{document}
\title{Quantum Interference Mechanism of Cooperative
Optical Phenomena in Extended Media}
\author{Valery I. Rupasov}
\address{Department of Physics, University of Toronto, Toronto,
Ontario, Canada M5S 1A7\\
and Landau Institute for Theoretical Physics, Moscow, Russia}
\date{\today}
\maketitle
\begin{abstract}
In the quantum process of stimulated Raman scattering
(SRS), a laser photon propagating in a resonance medium
undergoes multifold conversions into a Stokes photon
and back. The nontrivial ``cooperative'' behavior of
the Stokes component of light transmitted through the
medium is proven to be completely determined by the
interference of scattering amplitudes in different
sub-channels of the Stokes channel, which obviously
combines all the sub-channels with an odd number of
photon conversions. The theory of superfluorescence is
then derived as the limiting case of the SRS theory.
\end{abstract}


It has been typical to treat cooperative phenomena of
resonance quantum optics in an extended medium, such
as stimulated Raman scattering (SRS) \cite{SRS} and
superfluorescence (SF) \cite{SF}, by first describing
the initial growth of the Stokes pulse in SRS (or,
respectively, the radiation emitted in a SF process)
by linear quantum theory and then the succeeding evolution
by the classical nonlinear equations. It is assumed that
once the Stokes wave is initialized by the quantum vacuum
fluctuations and the Stokes pulse has grown to a classical
value, then the classical equations become valid. Both
the classical Maxwell-Bloch (MB) equations describing SF
\cite{MF} and a set of the classical equations describing
SRS \cite{DWC} were proven \cite{L,AKNS,AKN,K,K2,S,MM} to
be integrable. That allows one to apply the inverse scattering
method \cite{ISM} to study the time evolution of the SF
radiation and Stokes pulses within a medium.

In this Letter, we propose and discuss in detail a
microscopic mechanism of cooperative optical phenomena
based on studies of a quantum version of the standard
classical model of SRS.

Propagating in a partially excited resonance medium,
an incident laser photon undergoes multifold conversions
into a Stokes photon and back. The number of these
conversions defines a sub-channel of the scattering
process. A scattering amplitude in the elastic (Rayleigh)
channel is obviously given by a sum of scattering amplitudes
in sub-channels with an even number of conversions. In
this channel, an incident laser photon leaves a medium
preserving the frequency. Correspondingly, a scattering
amplitude in the inelastic (Stokes) channel is given by
a sum of the scattering amplitudes in sub-channels with
an odd number of conversions. Therefore, a photon leaves
a medium as a Stokes one.

Analyzing explicit expressions for the scattering
amplitudes, we show that a nontrivial ``cooperative''
behavior of the Stokes component of the radiation
transmitted through a medium is completely determined
by the interference of the scattering amplitudes in
different sub-channels of the inelastic channel of
scattering. The interference of the scattering amplitudes
in different sub-channel of the elastic channel results
correspondingly in the ``cooperative'' behavior of the
Rayleigh component of transmitted light.

A solution of the quantum model of SRS shows that there
is neither ``linear quantum stage'' nor ``succeeding
nonlinear classical evolution'' of the Raman process.
The quantum interference mechanism describes the
``cooperative'' character of SRS in a unified manner for
all incident photons. Moreover, the scattering amplitudes
do not depend on any physical parameters of an incident
laser pulse, such as its amplitude, duration, etc. They
are determined only by the quantum state of a medium.
Scattering in a medium, even in the elastic channel, an
incident laser photon changes the quantum state of a medium.
This results in changing the scattering amplitudes for all
successive photons, explaining thus the ``cooperative''
behavior of the transmitted light.

When the number of excitations of a medium grows, a number
of possible inelastic sub-channels also grows, and the
probability of scattering in the Stokes channel increases.
However, if a number of excited atoms exceeds a half of
the total number of atoms in a medium, further excitation
results in a decrease in the number of inelastic sub-channels.
Therefore, the probability of scattering in the Stokes channel
begins to fall, and vanishes asymptotically in the limit of
a completely inverted medium. The intensity of the Stokes
radiation coming out of a medium has a pulse (or soliton-like)
shape.

Finally, we show that a quantum model of SF in an extended
medium (or a quantum version of the classical MB model)
does not require a special consideration. Its solution
is derived as the limiting case of a solution of the
SRS problem with an appropriate initial condition.

The ``cooperative'' behavior of the radiation emitted by an
initially excited system of two-level atoms is determined by
quantum interference of amplitudes in different sub-channels
of emission. In the SF case, the sub-channels of the emission
process are characterized by different numbers of possible
reabsorption processes of a photon propagating in a partially
excited medium.

Integrability of the quantum model of SRS allows one to derive
an exact expression for the out-state of a scattering process
\cite{R}. Unfortunately, a complex algebra of operators involved
in this expression does not admit an analytical computation of
physical observables of the transmitted radiation. The problem
of a computation of physical observables is equivalent to the
problem of correlation functions in the theory of integrable
quantum systems \cite{KBI}. However, the analytical results
suggest a simple scheme for numerical calculations and numerical
simulations of the SRS process in terms of the six-vertex model
of statistical mechanics \cite{B}.

The Hamiltonian of a quantum version of the standard
SRS model can be written in the following form \cite{R}:
\begin{equation}
H=-\int_{-\infty }^{\infty} dx
\left\{i\bar{\epsilon}_\sigma(x)\frac{\partial}{\partial x}\epsilon_\sigma(x)
+J\sum_{a=1}^{M}\delta(x-x_a)
\bar{\epsilon}_\sigma(x)\left[\sigma^+_{\sigma\mu}r^-_a
+\sigma^-_{\sigma\mu}r^+_a\right]\epsilon_\mu(x)\right\},
\end{equation}
where the summation over repeated indexes is assumed.
In Eq. (1), the operators of the ``slow'' envelopes of
the laser (pump), $E_L(x)$, and Stokes, $E_S(x)$, components
of light are combined into the isotopic spinor
$$
\epsilon_\sigma=\left(
\begin{array}{c}
E_L\\E_S
\end{array}
\right),\;\;\;
\bar{\epsilon}_\sigma=(E_L^\dagger,E_S^\dagger)
$$
with the commutator
$[\epsilon_\sigma(x),\bar{\epsilon}_\mu(y)]=\delta_{\sigma\mu}\delta(x-y)$,
$\sigma,\mu=L,S$. To keep our terminology simple, we use hereafter
the term ``spin'' for the field's subindex $\sigma$. Thus, the
radiation field is described in terms of particles with spin up
(a laser photon) and down (a Stokes photon).

To account for a possible saturation of a resonance
transition of a medium, we treat the medium as a discrete
set of $M$ two-level atoms with the coordinates $x_a$,
$a=1,\ldots,M$, along the light propagation axis - the
$x$ axis. All the atoms are assumed to be positioned
on an interval of size $l$, $0<x_1<\ldots<x_M<l$.
As usual, they are described by the spin-$\frac{1}{2}$
operators ${\bf r}_a=(r^x_a,r^y_a,r^z_a)$,
$r^\pm=(r^x\pm ir^y)/2$. The Pauli matrices
$\sigma^{\pm}_{\sigma\mu}$ act in the spin space of the
radiation. The first term of the model Hamiltonian describes
a free propagation of the field components in the positive
direction of the sample axis, while the second one represents
a medium-field interaction with the coupling constant $J$,
the two-level atoms playing the role of ``spin impurities''.

In the classical limit and in the limit of a continuous
description of the medium, the equations of motion for
the dynamical variables of the field plus medium system
coincide with the equations of the classical SRS theory
\cite{SRS,SF,DWC}. In the quantum approach, integrability
of the SRS model is obvious, because the Hamiltonian
(1) coincides with the Hamiltonian of the Kondo model
with anisotropic exchange coupling whose integrability
has been proven by Wiegmann \cite{W}.

The scattering of the $j$-th particle on the $a$-th
impurity is described by the scattering matrix \cite{W,R}
\begin{equation}
S_{ja}=\frac{1}{2}(1\otimes 1+\sigma^z_j\otimes r^z_a)+
\frac{b}{2}(1\otimes 1-\sigma^z_j\otimes r^z_a)
+c\,(\sigma^+_j\otimes r^-_a+\sigma^-_j\otimes r^+_a),
\end{equation}
where $\otimes$ implies the tensor multiplication, and the
scattering amplitudes $b=\cos{J}$ and $c=i\sin{J}$ correspond
to the elastic (Rayleigh) and inelastic (Stokes) channels,
respectively. For what follows, it is convenient to use the
independence of the $S$-matrix of a particle momentum \cite{N}
and to formulate the SRS problem only in terms of spin
variables of particles and impurities.

The initial state of the spontaneous SRS problem is then
given by
\begin{equation}
|{\rm In}\rangle=|\bar{\Omega}_{\text{p}}\rangle
\otimes|\Omega_{\text{i}}\rangle,
\end{equation}
where the states of $N$ incident particles (all spins up)
and $M$ impurities (all spins down) are defined
as $\sigma^+_j|\bar{\Omega}_{\text{p}}\rangle=0$ and
$r^-_a|\Omega_{\text{i}}\rangle=0$. It is convenient to
represent the particle-impurity scattering matrix (2) as
a $2\times 2$ matrix acting in the spin space of a particle,
while its matrix elements contain the spin variables of an
impurity,
$$
S=\left[
\begin{array}{cc}
\frac{1}{2}[(1+b)+(1-b)r_a^z]&c\,r_a^-\\
c\,r_a^+&\frac{1}{2}[(1+b)-(1-b)r_a^z]
\end{array}\right].
$$

The subsequent scattering of the $j$-th particle on the
system of $M$ impurities is then described by the
ordered scalar product of matrices $S_{ja}$,
$$
L_j=S_{jM}\ldots S_{j1}=\left[
\begin{array}{cc}
A&B\\
C&D
\end{array}\right],
$$
where the operators $A,B,C$, and $D$ act in the spin space of
impurities. Acting on the initial state (3), the monodromy
matrix $L_j$ obviously reduces to the matrix
$$
U_j=A+C\otimes\sigma^-_j,
$$
where the first term corresponds to the elastic channel of
photon scattering on the atomic system, while the second one
represents the inelastic channel. In the latter, an incident
laser photon is converted into a Stokes one, and simultaneously
an excitation in the atomic system is created by the operator
$C$. The exact final state of the spontaneous SRS problem is
thus given by the ordered product of the matrices $U_j$,
\begin{equation}
|{\rm Out_N}\rangle=(A+C\otimes\sigma_N^-)\ldots(A+C\otimes\sigma_1^-)
|{\rm In}\rangle.
\end{equation}

Unlike the SRS problem in a pointlike geometry of the atomic
system \cite{R,R2,R3,R4,KKR}, the algebra of the elements of
the monodromy matrix $L$ in an extended medium is not closed.
Therefore, calculations of expectation values on the out-state
(4) run into great mathematical troubles. Here, we compute the
scattering amplitudes only for first two incident photons to
clarify a microscopic quantum mechanism of the SRS phenomenon.

The first photon ($j=1$) propagates in the unexcited atomic system
and the explicit expression for the out-state is easily found to be
\begin{equation}
|{\rm Out_1}\rangle=b^M|{\rm In}\rangle+
c\sum_{a=1}^{M}b^{(a-1)}r_a^+|\Omega_i\rangle
\otimes\sigma_1^-|\bar{\Omega}_p\rangle.
\end{equation}
Eq. (5) has a clear physical meaning. The first term corresponds
to the elastic channel with the scattering amplitude
${\cal A}^{(1)}_L=b^M$ and the probability $P^{(1)}_L=b^{2M}$.
The second term represents the inelastic channel, in which the
laser photon is converted into a Stokes photon, and simultaneously
an excitation in the atomic system is created. The probability
amplitude of finding the $a$-th atom in the excited state,
$\phi_a=cb^{(a-1)}$, is naturally treated as the wave function
of a one-particle medium excitation created by the transmitted
photon.

In this expression, the term $c=i\sin{J}$ is the probability
amplitude to excite the $a$-th atom, while
$b^{(a-1)}=(\cos{J})^{(a-1)}$ is the probability amplitude
that the scattering on the preceding $a-1$ atoms was elastic.
Since $J\ll 1$, the wave function of the excitation takes
an obvious exponential form
\begin{equation}
\phi(x)=iJ\exp{(-J^2\rho x/2)}.
\end{equation}
Here, aa atomic number $a$ is represented as $a=\rho x$,
where $\rho=M/l$ is the linear density of the number of
atoms, and $x$ is the atomic coordinate measured from
the left edge of the medium.

Let us consider now the scattering process of the second
incident laser photon ($j=2$). This photon propagates in
the medium whose quantum state has been prepared by the
first photon. The probability of the elastic scattering of
the second particle is obviously given by
\begin{equation}
P^{(2)}_L=\langle{\rm Out_1}|AA|{\rm Out_1}\rangle,
\end{equation}
and we have
\begin{equation}
A|{\rm Out}_1\rangle=b^Mb^M|{\rm In}\rangle+b^{(M-1)}c
\left[\sum_{a=1}^{M}b^{(a-1)}r_a^+
+c^2\sum_{a_1=1}^{M}\sum_{a_2=1}^{a_1-1}b^{(a_2-1)}r_{a_2}^+\right]
|\Omega_i\rangle\otimes\sigma_1^-|\bar{\Omega}_p\rangle.
\end{equation}
Here, the first term describes the elastic propagation of
the particle in the unexcited medium, while the second and
third terms correspond to the particle propagation in the
medium excited by the first incident photon. The second term
represents a sub-channel of the elastic channel of scattering
in which a particle does not change the direction of its spin
propagating within the medium as a laser photon,
$1\stackrel{L}{\rightarrow}M$. The third term represents the
other possible sub-channel of the elastic scattering in which
a particle changes the direction of its spin twice. Since one
of the atoms, say $a_1$, is excited, a particle can change a
spin direction first exciting an atom with a number $a_2<a_1$,
and then deexciting the atom $a_1$,
$1\stackrel{L}{\rightarrow}a_2\stackrel{S}{\rightarrow}a_1
\stackrel{L}{\rightarrow}M$. Thus, on an interval between atoms
$a_2$ and $a_1$ a particle propagates as a Stokes photon,
but comes out of the medium as a laser photon.

Despite the factor $c^2$ being very small, the contribution
of the third term to the probability $P^{(2)}_L$ is not
small compared to the contribution of the second one.
Moreover, omitting the third term, we immediately find
that $P^{(2)}_L>P^{(1)}_L$, and hence the probability of
the inelastic scattering falls with the growth of a photon
number, $P^{(2)}_S<P^{(1)}_S$.  The existence of two
sub-channels of scattering for the second incident particle
plays a crucial role for the nontrivial behavior of the
probabilities of scattering. The interference of the
scattering amplitudes in two possible sub-channels completely
determines the growth of the Stokes component. In the first
sub-channel, the excited state of the atomic system is not
changed, while in the second sub-channel an excitation of
the atomic system created by the first incident photon is
shifted to the left edge of the medium.

One obtains from Eqs. (7) and (8),
\begin{equation}
P^{(2)}_L=p^{2M}+\left[4+p+\frac{M|c|^2(M|c|^2-4)}{1-p^M}\right]
p^{(M-1)}(1-p^M),
\end{equation}
where $p=b^2$, and hence, in the limit $J\ll 1$,
\begin{equation}
\frac{P^{(2)}_L}{P^{(1)}_L}=1-2(MJ^2)^2+{\cal O}[(MJ^2)^4].
\end{equation}
Thus, the interference of the scattering amplitudes in
two sub-channels of the elastic channel of scattering
drastically change the behavior of the probability
$P^{(2)}_L$. Now, $P^{(2)}_L<P^{(1)}_L$, and hence, the
probability of conversion of the second laser photon into
the Stokes one, $P^{(2)}_S=1-P^{(2)}_L$, is bigger than
this probability for the first transmitted photon,
$P^{(1)}_S=1-P^{(1)}_L$, $P^{(2)}_S>P^{(1)}_S$.

The first derivative over time of the Stokes component
intensity on the edge of an transmitted pulse, is easily
found to be
\begin{equation}
\frac{d}{dt}I_S(t=0)=I_0\frac{P^{(2)}_S-P^{(1)}_S}{P^{(1)}_S}
\sim 2I_0(MJ^2)^2,
\end{equation}
where $I_0$ is the intensity of an incident laser pulse.
Thus, the intensity of the Stokes component at the initial
moment of time grows with the velocity proportional to
$M^2$; this is traditionally treated as a signature of the
cooperative behavior of the SRS process.

Let us consider now an initial state of the scattering
problem in which all incident particles are Stokes photons
(all particle spins down), while all atoms are excited
(all impurity spins up),
\begin{equation}
|{\rm In}\rangle=|\Omega_{\text{p}}\rangle
\otimes|\bar{\Omega}_{\text{i}}\rangle,
\end{equation}
where as $\sigma^-_j|\Omega_{\text{p}}\rangle=0$ and
$r^+_a|\bar{\Omega}_{\text{i}}\rangle=0$. Then, the
transmitted radiation contains obviously both the Stokes
and laser components. It is clear also that the scattering
process results in a deexcitation of atomic system.

Let us consider the case of a single excited atom.
The probability amplitude $A_{\text{ex}}(N)$ to find
the atom in the excited state after scattering $N$
incident laser photons is easily found from Eq. (2),
$A_{\text{ex}}(N)=(\cos{J})^N$. Taking into account
that $J\ll 1$, we obtain for a time evolution
of the probability amplitude
\begin{equation}
A_{\text{ex}}(t)=\left(1-\frac{J^2}{2}\right)^N
=\exp{\left(-\frac{1}{2}\gamma t\right)}.
\end{equation}
Here, $\gamma=J^2\rho$, and the number of scattered
photons at the moment of time $t$ is represented
as $N=\rho t$, where $\rho$ is the linear density
of incident photons.

As it should be expected, the excited atomic state
decays exponentially with the decay constant $\gamma$
to the ground state. If
\begin{equation}
\rho\to\infty,\, J\to 0,\,\, \text{but}\,\,J^2\rho=\gamma=\text{constant},
\end{equation}
as is already assumed in Eq. (13), an incident field
is obviously eliminated from consideration, and the
decay can be treated as spontaneous one, in which a
single ``Stokes'' photon is emitted. In this limit,
we can replace $JE_L\to\sqrt{\gamma}$ and $E_S(x)\to E(x)$
to reduce the Hamiltonian of the SRS problem (1) to
the standard Hamiltonian of the quantum Maxwell-Bloch
model,
\begin{equation}
H=-i\int dx E^\dagger(x)\frac{\partial}{\partial x}E(x)
-\sqrt{\gamma}\sum_{a}\int dx
\left[E^\dagger(x)r^-_a +r^+_aE(x)\right]\delta(x-x_a).
\end{equation}

An equivalence of these two models in the limit (14)
has been used \cite{PR} to study the SRS problem in
the framework of Maxwell-Bloch equation. It is clear
that a solution of the more general SRS problem can
be reduced to the SF problem only under some special
conditions \cite{PR}. However, a solution of the SF
problem, including statistical properties of emitted
photons \cite{R5}, is exactly derived from a solution
of the SRS problem in the limit (14).

I am thankful to S. John for stimulating discussions.

\end{document}